\documentclass[12pt]{article}

\usepackage{amsmath}
\usepackage{amsthm}
\usepackage{amsfonts}           
\usepackage{amssymb}            

\newlength{\xtra}
\usepackage[all]{xy}           

\newcommand{\Span}{\mathop{\mathrm{span}}\nolimits}
\newcommand{\eqdef}{\stackrel{\scriptscriptstyle\mathrm{def}}{=}}

\newcommand{\R}{\ensuremath{{\mathbb{R}}}}
\newcommand{\C}{\ensuremath{{\mathbb{C}}}}

\newcommand{\suhat}{\ensuremath{{\widehat{\mathfrak{su}}}}}
\newcommand{\sohat}{\ensuremath{{\widehat{\mathfrak{so}}}}}
\newcommand{\g}{\ensuremath{{\mathfrak{g}}}}

\newcommand{\ie}{{i.e.}}
\newcommand{\Hcal}{\ensuremath{{\cal H}}}

\newcommand{\NIM}{\ensuremath{{\cal N}}}

\newcommand{\Z}{\mathbb{Z}}

\newcommand{\RP}{\ensuremath{\mathop{\mathbb{R}{\rm P}}}\nolimits}

\newcommand{\Ad}{{\mathop{\rm Ad}\nolimits}}
\newcommand{\Ind}{\mathop{\rm Ind}\nolimits}
\newcommand{\Res}{\mathop{\rm Res}\nolimits}

\newcommand{\placeholder}{\ensuremath{\,-\,}}

\newcommand{\Dbrane}{D--brane}

\newcommand{\Ktheory}{K--theory}
\newcommand{\Ktheories}{K--theories}
\newcommand{\Kgroup}{K--group}
\newcommand{\Kgroups}{K--groups}

\newcommand{\Cech}{{\v{C}ech}}

\newcommand{\Kunneth}{K{\"u}nneth}
\newcommand{\MV}{Mayer--Vietoris}

\newcommand{\Ncal}{\mathcal{N}}

\newcommand{\Hom}{{\mathop{\rm Hom}\nolimits}}
\newcommand{\Tor}{{\mathop{\rm Tor}\nolimits}}
\newcommand{\Sym}{{\mathop{\rm Sym}\nolimits}}
\newcommand{\Id}{{\mathop{\rm Id}\nolimits}}

\newcommand{\tHop}[1]{\ensuremath{\vphantom{H}^{#1}\!H}}
\newcommand{\tH}{\ensuremath{\tHop{t}}}
\newcommand{\tKop}[1]{\ensuremath{\vphantom{K}^{#1}\!K}}
\newcommand{\tK}{\ensuremath{\tKop{t}}}
\newcommand{\ptset}{\ensuremath{\{\text{pt.}\}}}

\newcommand{\tRop}[1]{\ensuremath{\vphantom{R}^{#1}\!R}}

\newcommand{\GGtwist}{\ensuremath{(-)}}
\newcommand{\BStwist}{\ensuremath{(+)}}

\newenvironment{descriptionlist}{%
\begin{list}%
{}%
{}}%
{\end{list}}

{\everymath{\displaystyle\everymath{}}\array}%
{\endarray}

\newtheorem{theorem}{Theorem}

\begin{document}

\begin{titlepage}
  \begin{flushright}
    hep-th/0403287\\
    UPR-1075-T\\
    DESY-04-054
  \end{flushright}
  \vspace*{\stretch{5}}
  \begin{center}
    \LARGE
    Supersymmetric WZW models and \\ twisted K--theory of SO(3)    
  \end{center}
  \vspace*{\stretch{2}}
  \begin{center}
        \begin{center}
        {\normalsize Volker Braun$\,^\star$ and
          Sakura Sch\"afer-Nameki$\,^\sharp$}\\
        \bigskip\medskip
        {\it 
        $^\star$ David Rittenhouse Laboratory, 
        University of Pennsylvania\\
        209 S. 33$\strut^{\,rd}$ Street, Philadelphia, PA 19104, 
        United States}\\
        \texttt{vbraun@physics.upenn.edu}\\
        \bigskip
        {\it        
        $^\sharp$ II. Institut f\"ur Theoretische Physik, 
        University of Hamburg\\
        Luruper Chaussee 149, 22761 Hamburg, Germany}\\
        \texttt{S.Schafer-Nameki@damtp.cam.ac.uk} 
        \end{center}
  \end{center}
  \vspace*{\stretch{2}}
\begin{center}
{\bf Abstract}
\end{center}

\noindent
We present an encompassing treatment of \Dbrane{} charges in
supersymmetric $SO(3)$ WZW models. There are two distinct
supersymmetric CFTs at each even level: the standard bosonic $SO(3)$
modular invariant tensored with free fermions, as well as a novel
twisted model. We calculate the relevant twisted \Ktheories{} and find
complete agreement with the CFT analysis of D--brane charges. The
K--theoretical computation in particular elucidates some important
aspects of ${\cal N}=1$ supersymmetric WZW models on non-simply
connected Lie groups.

\vspace*{\stretch{10}}
\noindent
March 2004
\end{titlepage}

\newpage
 

\tableofcontents


\section{Introduction}
\label{sec:intro}

Much evidence has been provided recently in support of the conjecture
that charges of D--branes in string theory are measured by
K--theory~\cite{MooreMinasian, WittenDbranesKtheory}.  Exactly
solvable conformal field theory (CFT) backgrounds such as
Wess--Zumino--Witten (WZW) models and cosets thereof have proven to be
particularly fruitful grounds for testing this claim.

The CFT description of D--branes is relatively well under control and
the charges for ${\cal N}=1$ supersymmetric WZW models on
simply-connected Lie groups have been computed
in~\cite{FredenhagenSchomerus, Bouwknegt, GaGaTwisted} and for ${\cal
  N}=2$ coset models in~\cite{MMSI, LercheWalcher}. Due to the
non-trivial NSNS 3-form flux in these backgrounds, the main actors on
the other side of the conjecture are twisted K--theories. For
simply-connected group manifolds these have been obtained
in~\cite{FredenhagenSchomerus, MaldacenaMooreSeiberg, VolkerLieK} and
for ${\cal N}=2$ coset models in~\cite{SakuraEquivariant,
  Sakuratocome}, and shown to be in perfect agreement with the CFT
prediction of the charge groups --- as far as these are accessible.
One of the most useful tools in determining the twisted \Kgroups{},
which will also feature prominently in the present paper, is the
seminal work by Freed, Hopkins and Teleman
(FHT)~\cite{FreedVerlindeAlg, FreedICM, FHTcomplex, FHTintegral}
relating twisted equivariant K--theory to the representation theory of
loop groups (see also~\cite{Mickelsson, MickelssonInv}). This allows
to reduce many of the K--theoretical computations to algebraic
problems.

The presence of at least $\Ncal=1$ supersymmetry is vital for the
comparison with \Ktheory. The importance of fermions should not be
surprising as \Ktheory{} has deep ties with spinors and the Dirac
operator, and in string theory it is general lore that there are no
conserved \Dbrane{} charges in the bosonic string. The most
transparent justification\footnote{We thank G. Moore for pointing this
  out.} of this point is as follows: from a boundary field theory
point of view the charges are determined by boundary conditions modulo
RG-flows.  Thus, in order to obtain non-trivial charges or
equivalently non-trivial path components of the boundary theory, it is
necessary to project out the unit operator~\cite{MooreKphysics}.

Based on the charge relations derived by Fredenhagen and
Schomerus~\cite{FredenhagenSchomerus}, recently, Gaberdiel and
Gannon~\cite{GaGaSO} determined the charges of D--branes in WZW models
on non-simply connected group manifolds. The purpose of the present
paper is to compute the corresponding K--theories for the simplest
such group, $SO_3\eqdef SO(3)$.

There is one key subtlety in the case of non-simply
connected groups, that makes the computations slightly more cumbersome
(and thus more interesting) compared to the simply-connected case.
K--theoretically this can be phrased as follows: in addition to the
standard twisting in $H^3(G; \Z)$ there is an additional possibility
to twist with an element in $H^1(G; \Z_2)$. In the case of interest
to us, $H^1(SO_3; \Z_2)=\Z_2$, which can be interpreted as an
additional grading of the twisted K--theories. This additional choice
has a precise counterpart in the world-sheet description, where it
corresponds to different spin-structures for the fermions. In fact,
this interpretation is most apparent using the identification proven
by Atiyah and Hopkins~\cite{AtiyahHopkinsKpm} of
$H^1(X;\Z_2)$--twisted K--theory with the Hopkins K--theory
$K_{\pm}(X)$, which made its first appearances in the context of
D--brane charges in $(-1)^{F}$ orbifolds, where $F$ is the
(left-moving) space-time
fermion number, see
e.g.~\cite{WittenDbranesKtheory, SakuraM}.

In summary, we obtain the following picture: let $G$ be a non-simply
connected group, with universal cover $\widetilde{G}$ such that
$G=\widetilde{G}/\Z_2$. Then one has in general two ${\cal N}=1$
supersymmetric WZW models for $G$, corresponding to the choices of
twistings in $H^1(G; \Z_2)=\Z_2$. Equivalently, these choices
distinguish two modular invariants corresponding to the WZW model on
$G$, in the precise sense that they are obtained as simple current
extensions 
from the supersymmetric WZW model on $\widetilde{G}$, which differ by
the action of a $\Z_2$ simple current on the free fermion theory. In case the
latter is trivial the resulting model is simply the tensor product of
the bosonic WZW model on $G$ as of~\cite{GepnerWitten, Felderone,
  Feldertwo} with free fermions. This is the kind of model that is
relevant for the discussion in~\cite{GaGaSO}.  On the other hand if
the action on the fermions is non-trivial, the resulting modular
invariant does not factor into bosonic and fermionic parts, and has
not been discussed in the literature. We shall refer to these models
as {\it \GGtwist--twisted} and {\it \BStwist--twisted} supersymmetric
WZW models on $G$, respectively. Clearly, it would be interesting to
systematically explore these models further. This construction has
also interesting applications in finding new symmetry-breaking
boundary conditions, which we shall comment upon in our concluding
remarks.

The outline of this paper is as follows. Section 2 gives an overview
of the conformal field-theoretical aspects of the supersymmetric WZW
models on $SO_3$, in particular giving a detailed exposition of the
two different choices of spin structures, and the charge groups in
either model. The K--theory computation comprises the main body of the
paper, starting with a purely topological computation in section 3.
This is then refined using FHT-like methods in chapter 4, where we
provide a complete derivation of the twisted K--theories for both
types of twists in $H^1(SO_3; \Z_2)$. We conclude in section 5 and
discuss various directions in which the present work can be extended.


\section{Supersymmetric WZW models on $\mathbf{SO_3}$}
\label{sec:WZW}

\subsection{The level manifesto}
\label{sec:kmanifesto}

Let us begin by addressing the technical and subtle, but very crucial
issue of the level or equivalently, the twisting or equivalently, the
NSNS flux. Although we are really only interested in the $SO_3$
supersymmetric WZW model we are about to encounter various auxiliary
WZW models. In addition, in view of the K--theory computation, we wish
to use a meaningful notation for the levels, where precisely the
positive integers are allowed. We will denote them as follows:
\begin{descriptionlist}
\item[$k$] The level of the bosonic WZW model on $SO_3$, i.e. $k=0$ is
  the model with only one primary field, $k=1$ is the next smallest
  model and so on. This is the integer that classifies the $LSO_3$
  central extension.
\item[$\kappa$] The $H^3(SO_3;\Z)=\Z$ twist in the corresponding
  \Ktheory{}, equivalently the NSNS background flux.
\end{descriptionlist}
The difference between the level and the flux is a constant called the
adjoint shift, in our case (see section~\ref{sec:poincare})
\begin{equation}
  \kappa = k + 1
\end{equation}
Now we are really interested only in the supersymmetric WZW model, where
\begin{descriptionlist}
\item[$\kappa$] The level of the $\Ncal=1$ supersymmetric WZW model on
  $SO_3$, i.e. the central element in the super Kac--Moody algebra.
\end{descriptionlist}
As we will discuss in more detail later, this is always a $\Z_2$
orbifold of the $\Ncal=1$ supersymmetric WZW model on $SU_2$ at level
$2\kappa$. As is well-known~\cite{KazamaSuzuki}, the latter model is
isomorphic to a level-shifted bosonic $SU_2$ WZW model together with a free
fermion theory where
\begin{descriptionlist}
  \item[$2\kappa-2=2k$] The level of this bosonic $SU_2$ WZW model.
\end{descriptionlist}
Our use of $k$ vs. $\kappa$ is the standard notation to distinguish
between the supersymmetric and bosonic levels, respectively.  For the
comparison with the CFT computation we should also comment upon the
relation of our conventions to the ones chosen
in~\cite{GaGaSO,BordaloLens}, where the authors study only the bosonic
$SO_3$ WZW model and denote its level by $k\in 2\Z$.  (i.e. with a
spurious factor of $2$). Denote their $k$ by
$k_\text{GG}$, then the following conversion rules apply:
\begin{equation}
  \label{eq:GGconvert}
  k = \frac{k_\text{GG}}{2}
  \,,\qquad 
  \kappa = \frac{k_\text{GG}}{2}+1 \,.
\end{equation}


\subsection{Supersymmetric WZW models}


Our present objective is to study \Dbrane{} charges in $\Ncal=1$
supersymmetric WZW models on $SO_3$. The key ingredient for the
construction is to observe that $SO_3 = SU_2/\Z_2$, where the $\Z_2$
acts as the antipodal map. The bosonic $SO_3$ WZW model can therefore
be constructed as a simple-current extension of the diagonal
$\suhat(2)_{2k}$ theory~\cite{GepnerWitten, Felderone, Feldertwo},
where the order $2$ simple current acts on the integrable highest
weights $\Lambda = [2k-\lambda, \lambda]$ with $\lambda \in 0\dots
2k$, as
\begin{equation}
  \label{eq:sutwosc}
  J :\qquad  [2k-\lambda, \lambda] \rightarrow [\lambda, 2k-\lambda] 
\,  .
\end{equation}
The thereby resulting state space for the WZW model on $SO_3$ is
(see~\cite{GaGaSO}) for $k$ odd and even, respectively,
\begin{equation}
\begin{split}
  \label{eq:statespacesgaga} 
  k\in 2\Z_\geq+1: \quad \Hcal_{SO_3} =&
  \bigoplus_{n=0}^{k} \Hcal_{2n} \otimes \bar{\Hcal}_{2n} \oplus
  \bigoplus_{n=1}^{k} \Hcal_{2n-1} \otimes \bar{\Hcal}_{2k-2n+1} 
  \\
  k\in 2\Z_\geq: \quad \Hcal_{SO_3} =& 
  \bigoplus_{n=0}^{k/2-1}
  \left( \Hcal_{2n} \oplus \Hcal_{2k-2n} \right) \otimes
  \left(\bar{\Hcal}_{2n}\oplus \bar{\Hcal}_{2k-2n}\right) 
  \\
  & \qquad \oplus 2
  \left(\Hcal_{k} \otimes \bar{\Hcal}_{k}\right) 
  \,. 
\end{split}
\end{equation}
But the bosonic theory does not have any conserved \Dbrane{} charges
and is not interesting for our purposes. We want to study the
supersymmetric version hereof.

The supersymmetric $\suhat (2)$ model at level $2 \kappa$ has a
description in terms of the chiral algebra
\begin{equation}
  \label{eq:susysu}
  {\cal A} = \suhat(2)_{2k} \oplus\sohat(3)_1
  \,,
\end{equation}
where $\kappa= k+1$. The diagonal modular invariant for
eq.~\eqref{eq:susysu} is
\begin{equation}
\begin{split}
  \label{eq:susysutwo}
  \Hcal_\text{diag} =& 
  \left( \bigoplus_{\lambda} \Hcal_\lambda \otimes
    \bar\Hcal_{\lambda}  \right) \otimes 
   \left( \bigoplus_{l=0,1,2} \Hcal_l \otimes
    \bar\Hcal_l  \right) 
  \\
  =&~ \Hcal_{su(2)_{2k}} \otimes \Hcal_{F}
  \,.    
\end{split}
\end{equation}
In particular, one obtains a supersymmetric WZW model on $SO_3$ by
tensoring eq.~\eqref{eq:statespacesgaga} with the state space
$\Hcal_F$ of the $\sohat(3)_1$ free fermion theory. This is the model
studied in~\cite{GaGaSO}.

In the above construction, it was assumed that the simple current acts
only on the $\suhat(2)_{2k}$ part. However, one could also
contemplate the following construction of a supersymmetric $SO_3$ WZW
model: Extend by the order $2$ simple current $\langle J \oplus j
\rangle$, where the simple current $j$ acts on the $\sohat(3)_1$
weights by
\begin{equation}
  \label{eq:sothreesc}
  j :\qquad  [2-l, l] \rightarrow [l, 2-l]
  \,,\qquad 
  l=0,1,2
  \,.
\end{equation}
The currents $J$ and $j$ generate a simple current
group ${\cal G}=\Z_2 \oplus \Z_2$ for the theory
eq.~\eqref{eq:susysu}. We shall be interested in the following $\Z_2$
subgroups of ${\cal G}$:
\begin{equation}
\begin{split}
  \text{\GGtwist{} twist: }\qquad  & {\cal G}_{\GGtwist} = 
  \langle J\oplus \Id \rangle 
  \\
  \text{\BStwist{} twist: }\qquad  & {\cal G}_{\BStwist} = 
  \langle J\oplus j \rangle 
  \,.
\end{split}
\end{equation}
The corresponding simple current extensions of
eq.~\eqref{eq:susysutwo} will be denoted by \textit{\GGtwist--twisted
  model} and \textit{\BStwist--twisted model}, respectively (this
notation will be explained in section~\ref{sec:poincare}). In
particular, the \GGtwist--twisted model is the one discussed
in~\cite{GaGaSO}.

The state space for the \GGtwist--twisted model is given by 
\begin{equation}
  \Hcal_{SO_3,(-)}= \Hcal_{SO_3} \otimes \Hcal_F
  \,.
\end{equation}
The state space for the \BStwist--twisted model is straight forward to
obtain using simple-current techniques. As this will not be of main
concern to our discussion, we shall leave this for future work. 
 
Note that if one was to extend the theory with all of ${\cal G}$,
non-trivial discrete torsion~\cite{Vafa,BordaloWZWorbifold} is allowed
as $H^2\big(\Z_2\times \Z_2, U_1\big)=\Z_2$. We will not pursue these
here since they do not
correspond to a bona fide (non-orbifolded) WZW model on $SO_3$.


\subsection{D--brane charges}

In~\cite{GaGaSO} the charge groups for the D--branes in the
\GGtwist--twisted model were computed, and obtained to be
\begin{equation}
  {\cal K}_{SO_3, (-)}=
  \begin{cases}
    \Z_2 \oplus \Z_2 & k_{GG} \equiv 0\mod 4 \\
    \Z_4 & k_{GG} \equiv 2\mod 4\,.
  \end{cases}
\end{equation}
The NIM-reps $\NIM_{\mu \lambda}\strut^{\nu}$ for the \BStwist--twisted
model follow straight forwardly, thanks to known simple-current
technology. The derivation of the charges necessitates a
generalization of~\cite{FredenhagenSchomerus} to supersymmetric CFT,
where the fermions do not necessarily factor out and therefore the
NIM--reps do not separate into an affine and a free fermion part.  We
leave this for future discussions, see also~\cite{FredenhagenSO3}.

We shall for the present paper content ourselves with the following
heuristic derivation of the charge groups.  Geometrically, the two
choices of twist correspond to the following identifications in the
$\suhat(2)$ WZW model. The \GGtwist--twisted case corresponds to the
superposition of the brane with charge $q_\lambda$ with its image
under the antipodal map, \ie{} the brane of charge
$q_{2k-\lambda}$. So in this model one superposes
the brane with its anti-brane (see also~\cite{GaGaSO}) resulting in
\begin{equation}
  \label{eq:typeAtwist}
  \begin{split}
    \text{$(-)$ twist}:\qquad  
    q_{(-), \lambda}
    =&~
    q_{\lambda} + q_{2k-\lambda} = 
    (\lambda+1+2k-\lambda+1)q_0 
    = \\ =&~ 
    (2\kappa) q_0= 0
    \,, 
  \end{split}
\end{equation}
using $q_\lambda = (\lambda +1) q_0$. Thus these branes do not carry
any non-trivial charges. If $2|k$ then there is a brane invariant
under the antipodal map, yielding a $\Z_2$ charge.

The \BStwist--twist on the other hand corresponds to superposing the
$q_\lambda$-charged brane with the anti-brane of the brane with weight
$2k-\lambda$, wherefore
\begin{equation}
  \label{eq:typeBtwist}
\begin{split}
  \text{\BStwist{} twist} :\qquad  
  q_{\BStwist, \lambda} &= q_{\lambda} - q_{2k-\lambda} = 
  ( 2\lambda-2k )q_0\\     
            &=  (2\lambda +2) q_0 \,, 
\end{split}
\end{equation}
which implies that the corresponding charge group is
$\Z_{k+1}=\Z_\kappa$.  Furthermore, the brane with label $\lambda=k$
carries charge: identifying it with its image under the antipodal map
results in an unoriented world-volume, thus allowing for at most
2-torsion charges.  The K--theory computation below will confirm this.
Clearly there are no space-filling D3--branes~\cite{WittenBranesAdS}.
We should stress that a proper derivation of the charge relations in
supersymmetric theories should confirm this.


\section{Pure topology}
\label{sec:topo}

\subsection{Quick review of twisted cohomology}
\label{sec:review}

The archetypical example of a twisted cohomology theory is \Cech{}
cohomology for a nontrivial $\Z$ bundle, that is instead of taking
constant coefficients we take them to be only locally constant but
with a monodromy around some noncontractible loop in our space $X$. 
This obviously changes the cohomology groups, for example
$H^0(X;\Z)=\Z$ for any connected space (given by the constants, i.e.
sections of the trivial $\Z$ bundle) whereas the twisted cohomology
group is $\tH^0(X;\Z)=0$: There are no sections in a nontrivial $\Z$
bundle except the zero section.

Clearly, the possible twists in ordinary cohomology are defined by
specifying the monodromies around noncontractible loops, so by a map
$\pi_1(X)\to GL_1(\Z) =\Z_2$. But since the target is abelian such a
group homomorphism must factor through the abelianization
$\pi_1(X)/[-,-] = H_1(X;\Z)$. So the twist represents an element of
the dual of homology, i.e. of $H^1(X;\Z_2)$. Technically the choice of
twist always depends on the representative of the cohomology class,
but different representatives of the twist class lead to isomorphic
(albeit not canonically) twisted cohomology theories. We will ignore
this subtlety usually.

Since we will use them shortly let us compute the (twisted) cohomology
groups of $\RP^3$, say using CW-cohomology. The real projective
$3$--space has a cell decomposition into a single cell $c_i$ in
dimensions $i=0$ to $3$, and each cell $c_i$ is attached such that two
points of $\partial c_i$ are identified with one point in the lower
dimensional skeleton $\Sigma_{i-1}$. So the attaching maps would be
degree $2$ if it were not for the orientation: for example the $2$
endpoints of the interval $\partial c_1$ map to $\Sigma_0=c_0$ but with
opposite orientation, so they cancel. So the cohomology is the
homology of the cochain complex:
\begin{equation}
  \label{eq:CohomologyRP3}
  H^i(\RP^3;\Z) = 
  H_i\Big( 
    \xymatrix{ 
      0 \ar[r] &
      \underline{\Z} \ar[r]^-{0} &
      \Z \ar[r]^-{2} &
      \Z \ar[r]^-{0} &
      \Z \ar[r]^-{0} &
      0 }
    \Big)      
  =
  \begin{cases}
    \Z    & i=3 \\
    \Z_2  & i=2 \\
    0     & i=1 \\
    \Z    & i=0 \,.
  \end{cases}
\end{equation}
Now since $H^1(\RP^3;\Z_2)=\Z_2$ there is also a twisted cohomology,
which we will denote $\tHop{-}(\RP^3;\Z)$. The twisting effects the
orientations in the boundary maps $\partial c_i\to \Sigma_{i-1}$:
Where the two contributions in the untwisted case added up, they now
cancel and vice versa. So the twisted cohomology is 
\begin{equation}
  \label{eq:twistedCohomologyRP3}
  \tHop{-}^i(\RP^3;\Z) = 
  H_i\Big( 
    \xymatrix{ 
      0 \ar[r] &
      \underline{\Z} \ar[r]^-{2} &
      \Z \ar[r]^-{0} &
      \Z \ar[r]^-{2} &
      \Z \ar[r]^-{0} &
      0 }
    \Big)      
  =
  \begin{cases}
    \Z_2  & i=3 \\
    0     & i=2 \\
    \Z_2  & i=1 \\
    0     & i=0\,.
  \end{cases}
\end{equation}
Now let us turn towards \Ktheory{}. Here it turns out that (some of)
the possible twists are representing a class in $H^1(X;\Z_2)\oplus
H^3(X;\Z)$. The effect of the twist in $H^1(X;\Z_2)$ is again a
twisted identification as one goes around a noncontractible
loop: If $[E]-[F]$ is an element in the twisted \Ktheory{} then the
bundles $E$, $F$ are exchanged as one goes around a ``twist'' loop.

The $H^3(X;\Z)$ part of the twist class can be understood from the
transition function point of view (see e.g.~\cite{Aussies}). By a standard argument this
corresponds to a $U_1$ valued function $\varphi_{ijk}:U_i\cap U_j \cap
U_k \to U_1$ on each triple overlap. The transition functions
$g_{ij}:U_i\cap U_j \to GL$ of a twisted bundle then do not quite fit
together, but up to a phase factor:
\begin{equation}
  g_{ij} g_{jk} g_{ki} = \varphi_{ijk} 
  \qquad \text{on}~U_i\cap U_j \cap U_k \,.
\end{equation}
Again there is a subtlety here in that for non torsion twist classes
one cannot use finite dimensional bundles, but this technical problem
can be dealt with so we will ignore it in the following.


\subsection{Twisted \Ktheory{}}
\label{sec:twistedK}

For any generalized cohomology theory there is some
Atiyah--Hirzebruch--Whitehead spectral sequence relating it to
ordinary cohomology. For the case at hand this is the following:
\begin{theorem}[Generalized Rosenberg spectral sequence]
  Fix a closed manifold $X$ and let $t_1 \oplus t_3$ be a cocycle in
  $H^1(X;\Z_2)\oplus H^3(X;\Z)$. Then there is a $\Z\oplus \Z_2$
  graded spectral sequence with
  \begin{equation}
    \label{eq:RosenbergSS}
    E_2^{p,q} = 
    \tHop{t_1}^p\Big( 
      X;~ K^q(X) \Big) \,,
  \end{equation}
  converging to the twisted \Ktheory{} $\tKop{t_1\oplus t_3}^\ast(X)$.
  The spectral sequence is bounded in $p$ and moreover the first
  differential is $d_3 = \mathrm{Sq}_3 + t_3 \cup$.
\end{theorem}
\begin{proof}
  The only novelty is the $t_1\in H^1(X;\Z_2)$ twist, everything else
  can be found in~\cite{RosenbergSS}. Again let $X^n$ be the
  $n$--skeleton of a cell decomposition of $X$. Then there is a
  spectral sequence with 
  \begin{equation}
    E_1^{p,q} = 
    \tKop{t_1\oplus t_3}^{p+q}\Big(X^p,X^{p-1}\Big) \simeq
    K^{p+q}\Big(X^p,X^{p-1}\Big) \simeq
    K^q\big(\ptset\big)
  \end{equation}
  The only novelty is the differential $d_1$, which is now the
  $t_1$--twisted coboundary operator. Hence the $E_2$ tableau is
  eq.~\eqref{eq:RosenbergSS}. 
\end{proof}
We are interested in $SO(3)\simeq \RP^3$ with the possible\footnote{Of
  course we are only interested in positive levels.} twists
\begin{equation}
  \label{eq:twistclasses}
  (\pm,\kappa) 
  ~\in~ \Z_2 \oplus \Z_> 
  ~\subset~ H^1(\RP^3;\Z_2)\oplus H^3(\RP^3;\Z) \,.
\end{equation}
In the \BStwist{} case we find ($2$--periodic in $q$)
\begin{equation}
  \label{eq:TWISTEDRosenbergE2}
\begin{gathered}
  E_2^{p,q} = 
  \vcenter{
  \xymatrix@=0.2cm{
   {\scriptstyle q=2} & 
   \ar[ddrrr]^(0.8){d_3=\kappa} 
   \Z \strut& 0 \strut& \Z_2 \strut& \Z \strut\\
   {\scriptstyle q=1} & 
   0 \strut& 0 \strut& 0 \strut& 0  \strut\\
   {\scriptstyle q=0} & 
   \ar[]+/d 0.5cm/+/l 0.6cm/;[uu]+/d 0.5cm/+/l 0.6cm/+/u 0.8cm/  
   \ar[]+/d 0.5cm/+/l 0.6cm/;[rrr]+/d 0.5cm/+/l 0.6cm/+/r 1cm/
   \Z \strut& 0 \strut& \Z_2 \strut& \Z  \strut\\
   & {\scriptstyle p=0} 
   & {\scriptstyle p=1} 
   & {\scriptstyle p=2} 
   & {\scriptstyle p=3} 
  }};
  \quad
  E_3^{p,q} = 
  E_\infty^{p,q} = 
  \vcenter{
  \xymatrix@=0.2cm{
   {\scriptstyle q=2} & 
   0 \strut& 0 \strut& \Z_2 \strut& \Z_\kappa  \strut\\
   {\scriptstyle q=1} & 
   0 \strut& 0 \strut& 0 \strut& 0  \strut\\
   {\scriptstyle q=0} & 
   \ar[]+/d 0.5cm/+/l 0.6cm/;[uu]+/d 0.5cm/+/l 0.6cm/+/u 0.8cm/  
   \ar[]+/d 0.5cm/+/l 0.6cm/;[rrr]+/d 0.5cm/+/l 0.6cm/+/r 1cm/
   0 \strut& 0 \strut& \Z_2 \strut& \Z_\kappa  \strut\\
   & {\scriptstyle p=0} 
   & {\scriptstyle p=1} 
   & {\scriptstyle p=2} 
   & {\scriptstyle p=3} 
  }}
  \\[2ex]
  \Rightarrow \qquad 
  \tKop{(+,\kappa)}^1(\RP^3) = \Z_\kappa
  ,\quad
  \tKop{(+,\kappa)}^0(\RP^3) = \Z_2  \,,
\end{gathered}
\end{equation}
whereas in the \GGtwist{}--twisted case we obtain
\begin{equation}
  \label{eq:RosenbergE2}
\begin{gathered}
  E_2^{p,q} = 
  E_\infty^{p,q} = 
  \vcenter{
  \xymatrix@=0.2cm{
   {\scriptstyle q=2} & 
   0 \strut& \Z_2 \strut& 0 \strut& \Z_2  \strut\\
   {\scriptstyle q=1} & 
   0 \strut& 0 \strut& 0 \strut& 0  \strut\\
   {\scriptstyle q=0} & 
   \ar[]+/d 0.5cm/+/l 0.6cm/;[uu]+/d 0.5cm/+/l 0.6cm/+/u 0.8cm/  
   \ar[]+/d 0.5cm/+/l 0.6cm/;[rrr]+/d 0.5cm/+/l 0.6cm/+/r 1cm/
   0 \strut& \Z_2 \strut& 0 \strut& \Z_2  \strut\\
   & {\scriptstyle p=0} 
   & {\scriptstyle p=1} 
   & {\scriptstyle p=2} 
   & {\scriptstyle p=3} 
  }}
  \\[2ex]
  \Rightarrow \qquad 
  \tKop{(-,\kappa)}^1(\RP^3) = 
  \Z_2\oplus \Z_2 ~\text{or}~ \Z_4
  ,\quad
  \tKop{(-,\kappa)}^0(\RP^3) = 0 \,.
\end{gathered}
\end{equation}
Almost everything is determined directly from our knowledge of
Rosenberg's spectral sequence. We are left only with one tiny
ambiguity, we cannot decide the group law on the order $4$ charge
group $\tKop{(-,\kappa)}^1(\RP^3)$.

In general we expect for each possible twisted \Ktheory{} some CFT or
string compactification unless there is some physical reason why this
particular choice of discrete torsion is forbidden. So we should
expect there to be different WZW models for every choice of twist
$(\pm,\kappa) \in \Z_2 \oplus \Z_>$. Especially we should not be too
surprised if the order of the charge group is independent of the
level.

It remains of course to decide the final ambiguity, but this is
surprisingly hard compared to how easily we found almost the complete
answer. The actual computation will be in section~\ref{sec:FHT} and is
quite lengthy, but has the redeeming feature that it makes contact
with CFT methods.

For now let us have a closer look at the $(-,0)$--twisted \Ktheory{},
where we can use a simple trick to discern between the two
possibilities for $\tKop{(-,0)}^1(\RP^3)$. The idea (see
also~\cite{SakuraM} for a similar use) is that this \Ktheory{} is the
$K_\pm(S^3)$ of~\cite{AtiyahHopkinsKpm}, where $S^3$ comes with the
antipodal involution.  Then this \Kgroup{} can be computed as the
ordinary \Ktheory{} of $L\eqdef \big(S^{(4,0)}\times
R^{(1,1)}\big)/\Z_2$, the nontrivial real line bundle over $\RP^3$.
Now it happens that the one point compactification of $L$ is smooth,
and in fact $\RP^4$. Hence we find
\begin{equation}
\label{eq:tKRP3trick}
\begin{split}
  \tKop{(-,0)}^i(\RP^3) =&~ 
  K_{\pm}^i(S^3) = 
  K^{i+1}(L) = 
  K^{i+1}(\RP^4-\ptset) = \\ =&~
  \widetilde{K}^{i+1}(\RP^4) = 
  \begin{cases}
    \Z_4 & i=1 \\
    0    & i=0  \,.
  \end{cases}
\end{split}
\end{equation}
Unfortunately this trick is not easily extended to twistings
$(-,\kappa)$ with $\kappa>0$, but it is already tantalizing to see
that the twisted \Kgroups{} are indeed different from the twisted
cohomology $\tHop{-}^\ast(\RP^3)=\Z_2\oplus \Z_2$.


\section{FHT computation for $\mathbf{SO_3}$}
\label{sec:FHT}

So far we only used purely topological methods to find the relevant
\Kgroups. This seems to yield the correct result, although we are
unable to resolve the remaining $\Z_2\oplus \Z_2$ vs. $\Z_4$
ambiguity. It would be nice if we could resolve this, and even more
interestingly, if we could draw a parallel between the representation
theoretic argument on the CFT side and the \Ktheory{} computation.

We achieve this in the by now familiar way
(see~\cite{SakuraEquivariant}): Use some basic tricks to rewrite the
desired \Kgroup{} $\tK(SO_3)$ as equivariant \Ktheory{} of some
product space, and then use the equivariant \Kunneth{}
theorem~\cite{RosenbergSchochet} to relate that to tensor products
involving $\tK_G(G^\Ad)$. The latter is --- by the FHT
theorem~\cite{FreedVerlindeAlg, FHTcomplex, FHTintegral} --- the
Verlinde algebra, also known as the fusion ring. The computation of
the twisted \Ktheory{} thus boils down to simple algebra involving
fusion rings. We will not make use of the FHT theorem directly but
determine all necessary rings in the following directly.

In particular we use
\begin{equation}
  \tK^\ast\left(SO_3\right) = 
  \tK_{SU_2}^\ast\Big(SO_3^\Ad \times SU_2^\mathrm{L} \Big) \,,
\end{equation}
where the superscripts $\Ad$, $\mathrm{L}$ denote the $SU_2$
group\footnote{It is important to use $SU_2$ (as opposed to $SO_3$)
  equivariant \Ktheory{} since the \Kunneth{} theorem would not hold
  in the latter case: $SO_3$ is not a Hodgkin group.} action:
\textbf{Ad}joint and \textbf{L}eft multiplication. It turns out that
the twist class is only on the first factor, so the Cartesian product
really is a product, even considering the twist. Then we can apply the
equivariant \Kunneth{} theorem to the effect that we get a spectral
sequence
\begin{equation}
\label{eq:KunnethSS}
\begin{split}
  E_2^{\ast,\ast} &=
  \Tor^{\ast}_{R(SU_2)} \Big(
  \tK^\ast_{SU_2}(SO_3^\Ad),~\Z\Big) 
  = \\ &=
  \Tor^{\ast}_{R(SU_2)} \Big(
  \tK^\ast_{SU_2}(SO_3^\Ad),~
  \tK^\ast_{SU_2}(SU_2^{\mathrm{L}}) \Big) 
  \\
  & \qquad \Rightarrow~
  \tK_{SU_2}^\ast\Big(SO_3^\Ad \times SU_2^\mathrm{L} \Big)
  = \tK^\ast\left(SO_3\right) \,.
  \end{split}
\end{equation}
So now we first have to compute the twisted equivariant \Kgroup{}
$\tK_{SU_2}(SO_3)$, which will occupy sections~\ref{sec:poincare}
to~\ref{sec:quotient} (Note that this is similar, but not quite the
same as $\tK_{SO_3}(SO_3)$, which is computed in~\cite{FHTcomplex}).
Then we will evaluate the \Kunneth{} spectral sequence in
sections~\ref{sec:tensor} and~\ref{sec:higherTor}. Finally we compare
our result with the CFT analysis in section~\ref{sec:gaga}.


\subsection{Poincar\'e duality and adjoint shift}
\label{sec:poincare}

We want to compute the \Dbrane{} charge group of the $\Ncal=1$
supersymmetric WZW model at level $\kappa$, so what is the correct
level for the corresponding bosonic WZW model? A partial answer is
well-known for simply connected Lie groups, where the level of the
auxiliary bosonic WZW model is $\kappa-h^\vee$. But this shift (by the dual
Coxeter number in the simply connected case) is not the whole story.
Really we have to shift by the twist class induced via the adjoint
representation $Ad:G\to SO(\g)$ from the element
(cf.~\cite{FHTcomplex})
\begin{equation}
  (-,1,-)\in
  H^1_{SO}(SO;\Z_2)\oplus H^3_{SO}(SO;\Z)\simeq 
  \Z_2 \oplus \Big( \Z\oplus \Z_2\Big)  \,.
\end{equation}
In our case we want to use $G=SO_3$ and then the double cover
$\widetilde{SO}_3=SU_2$, i.e. pull back
\begin{equation}
  H^\ast_{SO}(SO) \longrightarrow 
  H^\ast_{SO_3}\big(SO_3\big) \longrightarrow
  H^\ast_{SU_2}\big(SO_3\big) \,.
\end{equation}
It is easy to see that the final adjoint shift is 
\begin{equation}
  1 \in H^3_{SU_2}\big(SO_3;\Z\big) \simeq \Z
  \quad \text{and} \quad
  - \in H^1_{SU_2}\big(SO_3;\Z_2\big) \simeq \Z_2 \,.
\end{equation}
An important point is that we also have to flip the sign, so the
straightforward bosonic $SO_3$ WZW model tensored with free fermions
corresponds to the $-\in H^1(SO_3;\Z_2)$ twisted \Ktheory{}.

Furthermore it will be more convenient to calculate the twisted
equivariant \Kgroups{} in K--homology in the following. Poincar\'e
duality (see~\cite{FHTcomplex}) relates this back to the K--cohomology
as of
\begin{equation}
  \label{eq:twistedKtheoryPD}
  \tKop{(\epsilon,\kappa)}^i_G(X) = 
  \tKop{(\epsilon,\kappa)}^G_{\dim(G)-i}(X) \,.
\end{equation}


\subsection{\MV{} sequence}
\label{sec:MV}

$SU_2$ acts by conjugation on $SO_3 = SU_2 / \pm$. There are three
kinds of orbits:
\begin{itemize}
\item The orbit of $1\in SU_2/\pm$, a fixed point. 
  \\
  The stabilizer is the whole $SU_2$.
\item The orbit of 
  $\left(\begin{smallmatrix} 0 & -1 \\ 1 & 0 
    \end{smallmatrix}\right) \in SU_2/\pm$, which is topologically
  $\RP^2$. 
  \\
  The stabilizer is $\Z_2 \ltimes U_1 \subset SU_2$.
\item The orbit of a generic point is an $S^2$.
  \\
  The stabilizer is a
  $U_1\subset SU_2$. 
\end{itemize}
This suggests the following cell decomposition of $SO_3 \simeq \RP^3$:
\begin{descriptionlist}
\item[$U$] The complement of the $\RP^2$.
\item[$V$] The complement of $1\in SU_2/\pm$.
\end{descriptionlist}
such that 
\begin{descriptionlist}
\item[$U$] is contractible to the fixed point.
\item[$V$] is contractible to the special $\RP^2$ orbit.
\item[$U\cap V$] is contractible to a generic $S^2$ orbit.
\end{descriptionlist}
The \MV{} sequence in K--homology for the cover $U$, $V$ of $SO_3$ is
then the following $6$ term cyclic exact sequence:
\begin{equation}
  \label{eq:6cyclicK}
  \vcenter{\xymatrix{
    \tK_1^{SU_2}(SO_3^\Ad)
    \ar[d]
    & 
    \ar[l]
    \tK_1^{SU_2}(\RP^2)
    & 
    \ar[l]
    0
    \\
    R(U_1)
    \ar[r]^-{\Ind}
    & 
    R(SU_2)
    \oplus
    \tK_0^{SU_2}(\RP^2)
    \ar[r]
    & 
    \tK_0^{SU_2}(SO_3^\Ad)\,.
    \ar[u]
  }}
\end{equation}
The inclusion $i:\RP^2\hookrightarrow \RP^3$ identifies the cohomology groups
\begin{equation}
  i^\ast: H^1(\RP^3;\Z_2) 
  \stackrel{\sim}{\longrightarrow} 
  H^1(\RP^2;\Z_2) 
  \,,
\end{equation}
so we can take the cocycle's support disjoint from $U$ in the cyclic
exact sequence.

Now concerning the \Kgroups{} of $\RP^2$, they are again the
representation ring of the stabilizer, as $SU_2$ acts transitively.
But there is a subtlety as this cell might come with a nontrivial
twist class. Even more delicately, the identification of the
K--homology groups with the representation ring uses Poincar\'e
duality, and this flips the sign of the twist as $\RP^2$ is not
orientable:
\begin{equation}
  \label{eq:twistedpoincare}
  \tKop{t}_0^{SU_2}\left(\RP^2\right) = 
  \tKop{-t}^0_{SU_2}\left(\RP^2\right) =
  \tRop{-t}\left( \Z_2 \ltimes U_1 \right)
  \,.
\end{equation}


\subsection{The (twisted) representation rings}
\label{sec:RG}

Let us review the representation rings that occur in our discussion to
fix notation. The most important one is for $SU_2$, since everything
in eq.~\eqref{eq:6cyclicK} is an $R(SU_2)$ module:
\begin{equation}
  \label{eq:RSU2}
  R(SU_2) = \Z[ \Lambda ] \,,
\end{equation}
generated by the fundamental ($2$ dimensional) representation
$\Lambda$. Instead of taking powers of $\Lambda$ there is a different
$\Z$ basis that is very useful in practice. This basis are the
irreducible representations of $SU_2$, which are all symmetric powers
of $\Lambda$. They are given recursively as
\begin{align}
  \label{eq:SymRecursive}
  \Sym^{-1}(\Lambda) =&~ 0
  \notag \\
  \Sym^{0}(\Lambda) =&~ 1
  \notag \\
  \Lambda\, \Sym^n{\Lambda} =&~ 
  \Sym^{n+1}(\Lambda) + \Sym^{n-1}(\Lambda) \,.
\end{align}
Next we have the representations of $U_1$, those are
\begin{equation}
  \label{eq:RU1}
  R(U_1) = \Z[\alpha, \alpha^{-1} ] \,,
\end{equation}
with $R(SU_2)$ module structure $\mu: R(SU_2)\times R(U_1) \to R(U_1)$
induced by the embedding $U_1 \subset SU_2$. Explicitly the $R(SU_2)$
action is given by
\begin{equation}
  \label{eq:RU1module}
  \mu(\Lambda, x) = (\alpha + \alpha^{-1}) x 
  \quad\forall x\in R(U_1)  
  \,.
\end{equation}

The representation theory of the semidirect product $\Z_2\ltimes U_1$
is more complicated. This is abstractly the group
\begin{equation}
  \label{eq:Z2sdpU1}
  \Z_2\ltimes U_1 = 
  \big\{ (s,\phi): s\in \Z_2,~\phi\in \R/2\pi \Z \big\}
  , \quad
  (s_1,\phi_1)\cdot (s_2, \phi_2) = 
  (s_1 s_2, \phi_1 + s_1 \phi_2 ) \,,
\end{equation}
which is the same as $O_2$, but more naturally we should think of it
as the double cover of $O_2$.  There are two obvious one dimensional
representations, the trivial and the sign representation $\sigma$. In
addition to those we also have the $2$ dimensional representation from
the embedding $\Z_2\ltimes U_1 \subset SU_2$, which we call again
$\Lambda$ (this notation makes sense, since by definition then the
$R(SU_2)$ module structure is just multiplication)
\begin{equation}
  \label{eq:sdp2Drep}
  \Lambda(+1,\phi) = 
  \begin{pmatrix}
    e^{-i\phi} & 0 \\ 0 & e^{i \phi}
  \end{pmatrix}
  \quad
  \Lambda(-1,\psi) = 
  \begin{pmatrix}
    0 & -e^{i \psi} \\ e^{i \psi} & 0
  \end{pmatrix} \,.
\end{equation}
One can easily check that $\Lambda\otimes \sigma$ is conjugate to
$\Lambda\otimes 1 = \Lambda$, while $\sigma$ is of course not
conjugate to $1$. Hence the representation ring is
\begin{equation}
  \label{eq:RZ2sdpU1}
  \tRop{+}(\Z_2 \ltimes U_1) \eqdef
  R(\Z_2 \ltimes U_1) =
  \Z[\Lambda, \sigma] 
  \Big/ 
  \left\langle
  \sigma^2-1,~
  \Lambda(\sigma - 1)
  \right\rangle \,.
\end{equation}
Finally, there is the possibility to twist the $\Z_2\ltimes U_1$
representations, and we get the corresponding twisted representation
rings. Really those are defined as the twisted equivariant \Kgroups{}
of a point, and for the case at hand are (see~\cite{FreedICM}):
\begin{subequations}
\begin{align}
  \label{eq:tR1sdp}
  \tKop{-}^1_{\Z_2 \ltimes U_1}\big( \ptset \big) \eqdef &~
  \tRop{-}^1\big(\Z_2 \ltimes U_1\big) =
  \left\langle \sigma-1\right\rangle_{R(\Z_2 \ltimes U_1)}
  \\ 
  \label{eq:tR2sdp}
  \tKop{-}^0_{\Z_2 \ltimes U_1}\big( \ptset \big) \eqdef&~
  \tRop{-}\big(\Z_2 \ltimes U_1\big) =
  \left\langle \sigma+1\right\rangle_{R(\Z_2 \ltimes U_1)}
  \\ \notag   
  \simeq&~ R(SU_2) =\Z[\Lambda]
  ~\text{as $R(SU_2)$ module} \,.
\end{align}
\end{subequations}


\subsection{Dirac induction}
\label{sec:dirac}

The essential part of the whole computation is to identify the map
dubbed $\Ind$ in eq.~\eqref{eq:6cyclicK}. The pushforward in
K--homology is actually Dirac induction, a version of Borel--Weil
induction that does not require complex structures
(see~\cite{FHTintegral})
\begin{equation}
  \label{eq:ind}
  \Ind: R(U_1) \to R(SU_2) \oplus \tRop{\pm}(\Z_2\ltimes U_1) \,.
\end{equation}
The first component is just the usual induction, precomposed with
multiplication by $\alpha^{\kappa}$ which is the effect of the twist
class $\kappa = k+1 \in H^3_{SU_2}(SO_3)$:
\begin{equation}
  \label{eq:Ind1}
  \pi_1\circ \Ind (\alpha^n) = 
  \Sym^{n+\kappa} \Lambda \,.
\end{equation}
Concerning the second component we have to distinguish between the
possible $\pm$ twists (for representation rings it makes also sense to
call this a grading), we will come to that shortly.

Having identified the induction map and assuming that $\kappa>0$ (as
we will always do) it is then easy to see that the total $\Ind$ is
injective, so we can indeed determine all the unknowns in the exact
sequence eq.~\eqref{eq:6cyclicK}:
\begin{subequations}
\begin{align}
  \label{eq:tK1resultInd}  
  \tK^{SU_2}_1\big(SO_3\big) &=\, 
  \tRop{-t}^1\big(\Z_2 \ltimes U_1\big)
  \\
  \label{eq:tK0resultInd}  
  \tK^{SU_2}_0\big(SO_3\big) &=\,
  \Big(
  R(SU_2) \oplus 
  \tRop{-t}\big(\Z_2 \ltimes U_1\big)  
  \Big)
  \Big/ \Ind\big( R(U_1) \big)\,.
\end{align}
\end{subequations}
Moreover the first component turns out to be surjective, so we can
write $\tK^{SU_2}_0\big(SO_3\big)$ as a quotient of
$\tRop{-t}\big(\Z_2 \ltimes U_1\big)$ only.

\subsubsection{Twisted Poincar\'e duality}

It remains to identify the second component of the induction map
eq.~\eqref{eq:ind}. This is almost, but not quite, the Dirac induction
\begin{equation}
  \label{eq:indpm}
  \Ind^\pm: R(U_1) \to \tRop{\pm}(\Z_2\ltimes U_1) \,.
\end{equation}
An important subtlety here is that in the twisted Poincar\'e duality
$\tKop{\pm}^0_{SU_2}\left(\RP^2\right) =
\tKop{\mp}_0^{SU_2}\left(\RP^2\right)$ we had to pick a fundamental
class, or dually a class in $\tKop{-}^0_{SU_2}\left(\RP^2\right)$.
But the generator is $1\in \tRop{-}(\Z_2\ltimes U_1)$ which is a
nontrivial\footnote{In other words, the trivial representation is not
  ``$-$'' twisted.} twisted representation.

To compensate for this we have to make sure that our K-homology
pushforward $\pi_2 \circ \Ind(1)$ is again the fundamental
class. From that we can identify\footnote{Or
  $\Ind^\pm(\alpha^{-1}\cdot \placeholder)$, depending on the chosen
  orientation.} 
\begin{equation}
  \label{eq:twistedpoincareInd}
  \pi_2 \circ \Ind( \placeholder ) = 
  \Ind^\pm( \alpha \cdot \placeholder) \,,
\end{equation}
using the results on $\Ind^\pm$ from the remainder of this section.

\subsubsection{Ungraded Induction}

First, let us look at the induction involving only untwisted
representation rings (this computes then the \GGtwist--twisted
\Ktheory). We find 
\begin{subequations}
\begin{align}
  \label{eq:Ind2untwisted1}
  \pi_2 \circ \Ind(1) =
  \Ind^+(\alpha)
  =&~ \Lambda
  \\ 
  \label{eq:Ind2untwisted2}
  \pi_2 \circ \Ind(\alpha^{-1}) =
  \Ind^+(1) 
  =&~ 1+\sigma    \,.
\end{align}
\end{subequations}
Let us pause to explain the latter eq.~\eqref{eq:Ind2untwisted2},
which might be less obvious. This is a rather degenerate case of Dirac
induction as the quotient $(\Z_2\ltimes U_1)\big/U_1 \simeq \Z_2$ is
$0$--dimensional.

Recall the usual Dirac induction for a subgroup $H\subset G$, see
e.g.~\cite{Sternberg,Landweber}: Given a representation $\rho: H\to V$
we can construct a $G$ representation on $\Gamma(G\times_H V)$. The
problem is that the latter (the space of sections) will in general not
be finite dimensional. The solution is to define an elliptic operator
$D:\Gamma(G\times_H V_1)\to \Gamma(G\times_H V_2)$, then the $G$
equivariant index $\mathrm{Index}_G(D) \in R(G)$ yields a finite
dimensional (virtual) representation.

But in the case at hand $G/H \simeq \Z_2$ is just two points, so
$\Gamma(G\times_H V)$ is $2 \dim(V)$ dimensional. Especially for
$V=\C$ the trivial representation we see that 
$\Gamma(G\times_H
\C)=L^2(\Z_2)=\Span_\C(1, \sigma)$ is generated by the trivial and
the sign representation, this explains eq.~\eqref{eq:Ind2untwisted2}.

\subsubsection{Graded Induction}

The induction to $\tRop{-}\big(\Z_2 \ltimes U_1\big)$ (which necessary
for the $(+)$--twisted \Ktheory) is related to
the untwisted restriction and induction via the following diagram with
exact rows
\begin{equation}
  \label{eq:RGses}
  \vcenter{\xymatrix{
    R\big(\Z_2 \ltimes U_1\big) \ar[r]^-{\Res^{+}} & 
    R\big(U_1\big) \ar[r]^-{\Ind^{-}} \ar@{=}[d] & 
    \tRop{-}\big(\Z_2 \ltimes U_1\big) \\
    R\big(\Z_2 \ltimes U_1\big)  & 
    R\big(U_1\big) \ar[l]_-{\Ind^{+}} & 
    \tRop{-}\big(\Z_2 \ltimes U_1\big) \ar[l]_-{\Res^{-}} \,,
  }}
\end{equation}
and moreover induction and restriction are adjoint functors, i.e.
\begin{equation}
  \label{eq:adjoint}
  \Hom_{R(U_1)}( \Res^\pm \placeholder , \placeholder ) 
  =
  \Hom_{\tRop{\pm}(\Z_2 \ltimes U_1)}
  ( \placeholder , \Ind^\pm \placeholder)   \,.
\end{equation}
Furthermore all maps are $R\big(\Z_2 \ltimes U_1\big)$--module maps
using the module structure discussed in section~\ref{sec:RG}, so to
specify the maps we just have to write down the image of generators in
the respective presentations
eqns.~\eqref{eq:RZ2sdpU1},\eqref{eq:tR2sdp},\eqref{eq:RU1}:
\begin{equation}
  \label{eq:RGpmIndRes}
  \begin{split}
    &\Res^{+}\left(1\right) = 1
    \\
    &\Res^{-}\left(1\right) = \alpha-\alpha^{-1} \,.    
  \end{split}
\end{equation}
The ordinary restriction $\Res^+$ is obvious. For the twisted
restriction $\Res^-$ note that $\alpha-\alpha^{-1}$ generates the
kernel of $\Ind^+$, and since the horizontal lines in
eq.~\eqref{eq:RGses} are exact this already fixes the
restriction\footnote{One can make this more precise using the description
  as super-representations.} (up to an irrelevant overall sign).

Now $\Ind^-$ is right adjoint to $\Res^-$, so e.g.
\begin{equation}
  \begin{split}
    \C \simeq
    \Hom_{R(U_1)}\Big( \alpha-\alpha^{-1} ,~ \alpha \Big) 
    =&~
    \Hom_{R(U_1)}\Big( \Res^- (1) ,~ \alpha \Big) =
    \\
    =&~
    \Hom_{\tRop{-}(\Z_2 \ltimes U_1)}
    \Big( 1 ,~ \Ind^- (\alpha) \Big) \,.    
  \end{split}
\end{equation}
Together with exactness of the top row in eq.~\eqref{eq:RGses} this
determines 
\begin{equation}
  \label{eq:Indminus}
  \begin{split}
    \pi_2 \circ \Ind(1)
    =&~
    \Ind^-(\alpha)=1
    \\
    \pi_2 \circ \Ind(\alpha^{-1})
    =&~
    \Ind^-(1) = 0 \,.
  \end{split}  
\end{equation}


\subsection{Determining the quotient}
\label{sec:quotient}

The representation ring $R(U_1)$ is a free $R(SU_2)$ module, generated
by $\alpha^n$ and $\alpha^{n+1}$ (i.e. any two consecutive powers of
$\alpha$ are $R(SU_2)$--linearly independent and generate all of
$R(U_1)$). Their image under the pushforward then generates
$\Ind\big(R(U_1)\big)$ as an $R(SU_2)$ module. Taking $n=-\kappa-1$,
the pushforward has the form
\begin{equation}
  \label{eq:IndImgGen}
  \begin{split}
    \Ind(\alpha^{-\kappa}) &=\, \Sym^0(\Lambda) \oplus \big(\cdots\big)
    = 1 \oplus \big(\cdots\big)
    \\
    \Ind(\alpha^{-\kappa-1}) &=\, \Sym^{-1}(\Lambda) \oplus
    \big(\cdots\big) = 0 \oplus \big(\pi_2 \circ
    \Ind(\alpha^{-\kappa-1})\big) \,.
  \end{split}
\end{equation}
So using the first relation we can write every equivalence
class in the quotient eq.~\eqref{eq:tK0resultInd} uniquely 
as $0\oplus
(\text{something})$. The second relation keeps that choice of
representative, so
\begin{equation}
  \label{eq:K0quotient}
  \tK^{SU_2}_0\big(SO_3\big) =
  \tRop{-t}\big(\Z_2 \ltimes U_1\big)  
  \Big/ \Big( \big( \pi_2 \circ
    \Ind(\alpha^{-\kappa-1})\big)  \cdot R(SU_2) \Big)  \,.
\end{equation}
In the \BStwist{} case we found
\begin{equation}
  \label{eq:Indalphatwisted}
  \pi_2 \circ \Ind(\alpha^n) = \Sym^n(\Lambda) 
  \quad \in \tRop{-}\big(\Z_2 \ltimes U_1\big) \,,
\end{equation}
in the same way as for the first component of $\Ind$. Applying
eq.~\eqref{eq:SymRecursive} we find that 
\begin{equation}
  \label{eq:Indalphatwistedfinal}
  \pi_2 \circ \Ind(\alpha^{-\kappa-1}) = 
  - \Sym^{\kappa-1}(\Lambda) =
  - \Lambda^{\kappa-1}+\cdots 
  \quad \in \tRop{-}\big(\Z_2 \ltimes U_1\big) \,,
\end{equation}
generates the relation.

In the \GGtwist{} case it is not quite so easy to write down a formula,
however we can find $\Ind(\alpha^n)$ recursively using the $R(SU_2)$
module structure
\begin{equation}
  \label{eq:IndRecursive}
  \Lambda \Ind(\alpha^{n}) = 
  \Ind \Big( \Lambda \alpha^{n} \Big) = 
  \Ind \Big( \big(\alpha+\alpha^{-1}\big) \alpha^{n} \Big) = 
  \Ind(\alpha^{n+1}) + \Ind(\alpha^{n-1}) \,,
\end{equation}
and eq.~\eqref{eq:Ind2untwisted1},\eqref{eq:Ind2untwisted2}. The
result is that
\begin{equation}
  \label{eq:Indalphauntwisted}
  \pi_2 \circ \Ind(\alpha^{-n-1}) = 
  \begin{cases}
    p_n(\Lambda) 
    & \forall n~\text{even} \\ 
    p_n(\Lambda) + (-1)^{\frac{n}{2}}(1+\sigma) 
    &  \forall n~\text{odd}
  \end{cases}
  \quad \in \tRop{+}\big(\Z_2 \ltimes U_1\big) \,,
\end{equation}
where $p_n$ is a polynomial of degree $|n|$ without constant part. Its
value at $2$ will be important in the following, by straightforward
induction one can show that
\begin{equation}
  \label{eq:pnEval2}
  p_n(2) = 
  \begin{cases}
    2 & \forall \,n\in 2\Z+1 \\
    4 & \forall \,n\in 4\Z+2 \\
    0 & \forall \,n\in 4\Z \,.
  \end{cases}
\end{equation}
Putting everything together we find\footnote{In the \GGtwist--twisted
  case here the $\langle\cdot\rangle$ means: Act with all of
  $\Z[\Lambda,\sigma]$. But we really only want to mod out the
  $R(SU_2)=\Z[\Lambda]$ image of $\Ind(\alpha^{-\kappa-1})$. However,
  this is the same thing since $\sigma p_n(\Lambda)=p_n(\Lambda)$,
  because $p_n$ does not have a constant part.}
\begin{equation}
  \label{eq:tKSU2SO3result}
  \begin{split}
    \tKop{(+,\kappa)}^{SU_2}_1\big(SO_3\big) &=\,
    \Z[\Lambda] / \Lambda
    \\
    \tKop{(+,\kappa)}^{SU_2}_0\big(SO_3\big) &=\,
    \Z[\Lambda] \Big/\! 
    \left\langle \Sym^{\kappa-1}(\Lambda)\right\rangle
    \\
    \tKop{(-,\kappa)}^{SU_2}_1\big(SO_3\big) &=\, 0
    \\
    \tKop{(-,\kappa~\text{odd})}^{SU_2}_0\big(SO_3\big) &=\,
    \Z[\Lambda,\sigma] \Big/\!  
    \left\langle 
      \Lambda(\sigma -1),~
      \sigma^2-1,~
      p_{\kappa}(\Lambda)\right\rangle
    \\
    \tKop{(-,\kappa~\text{even})}^{SU_2}_0\big(SO_3\big)  &=\,
    \Z[\Lambda,\sigma] \Big/\!  
    \left\langle 
      \Lambda (\sigma -1),~
      \sigma^2-1,~
      p_{\kappa}(\Lambda) + (-1)^{\frac{\kappa}{2}}(1+\sigma) 
    \right\rangle  
  \end{split}
\end{equation}
as $R(SU_2)=\Z[\Lambda]$ modules, the corresponding cohomology groups
are then determined by Poincar\'e duality
eq.~\eqref{eq:twistedKtheoryPD}.


\subsection{The tensor product}
\label{sec:tensor}

Now that we determined the twisted equivariant \Kgroups{} we can apply
the equivariant \Kunneth{} theorem and determine the corresponding
non--equivariant \Ktheory{}. The result is a spectral sequence with
\begin{equation}
  E_2^{\ast,\ast} = 
  \Tor_{R(SU_2)}^{\ast}\Big( \tK^\ast_{SU_2}(SO_3),~\Z \Big) \,,
\end{equation}
where $R(SU_2)=\Z[\Lambda]$ acts on $\Z$ by multiplication with the
dimension of the representation. With other words,
$-\otimes_{R(SU_2)}\Z$ is evaluation at $\Lambda=2$:
\begin{equation}
  \label{eq:tKSU2SO3tensorresult}
  \begin{split}
    \tKop{(+,\kappa)}^{SU_2}_1\big(SO_3\big) 
    \otimes_{R(SU_2)} \Z ~&=~
    \Z_2
    \\
    \tKop{(+,\kappa)}^{SU_2}_0\big(SO_3\big)  
    \otimes_{R(SU_2)} \Z ~&=~
    \Z_\kappa
    \\
    \tKop{(-,\kappa)}^{SU_2}_1\big(SO_3\big)
    \otimes_{R(SU_2)} \Z ~&=~ 0
    \\
    \tKop{(-,\kappa~\text{odd})}^{SU_2}_0\big(SO_3\big)
    \otimes_{R(SU_2)} \Z ~&=~
    \Z[\sigma] \Big/\!  
    \left\langle 
      2(\sigma -1),~
      \sigma^2-1,~
      2 \right\rangle 
    = \\ &=~
    \Z_2[\sigma]\Big/\! \langle \sigma^2-1 \rangle = \Z_2 \oplus \Z_2
    \\
    \tKop{(-,\kappa\in 4\Z_>)}^{SU_2}_0\big(SO_3\big)
    \otimes_{R(SU_2)} \Z ~&=~    
    \Z[\sigma] \Big/\!  
    \left\langle 
      2(\sigma -1),~
      \sigma^2-1,~
      1+\sigma
    \right\rangle
    = \Z_4
    \\
    \tKop{(-,\kappa\in 4\Z_\geq+2)}^{SU_2}_0\big(SO_3\big)
    \otimes_{R(SU_2)} \Z ~&=~
    \Z[\sigma] \Big/\!
    \left\langle 
      2(\sigma -1),~
      \sigma^2-1,~
      3-\sigma) 
    \right\rangle 
    = \\ &=~
    \Z \big/ \!\gcd( 4, 8 ) = \Z_4
    \,.
  \end{split}
\end{equation}


\subsection{Higher Tor}
\label{sec:higherTor}

The CFT charge equation~\cite{FredenhagenSchomerus}
\begin{equation}
  \label{eq:charge}
  \dim(\lambda) q_a =
  \sum \mathcal{N}_{\lambda a}\strut^{b} q_b \,,
\end{equation}
really tells you that the charge group is the tensor product
\begin{equation}
  \label{eq:Ntensor}
  \mathcal{N} \otimes_{RG} \Z \,,
\end{equation}
where $\mathcal{N}$ is the algebra of the $q_a$ with structure
constants $\mathcal{N}_{ab}\strut^c$.

But the derivation of the charge equation is by no means
mathematically strict. Indeed we know examples where the twisted
\Ktheory{} and hence the charge group is strictly bigger than
eq.~\eqref{eq:Ntensor}, for example most\footnote{With the exception
  of $SU_2$ at arbitrary level and other Lie groups at special levels
  where the \Kgroups{} vanish.} WZW models on compact simply connected
simple Lie groups (see~\cite{VolkerLieK}).  But by a generalized
nonsense argument involving the \Kunneth{} spectral sequence we know
that the tensor product eq.~\eqref{eq:Ntensor} is a subgroup of the
\Kgroup, i.e. there are no additional relations between the charges in
eq.~\eqref{eq:charge}.

But to find the whole charge group we must determine the whole
$\Tor(-,-)$, not just its degree zero piece $-\otimes-$. Since the
second argument $\Z=\Z[\Lambda]/\langle\Lambda-2\rangle$ has only one
relation as $R(SU_2)$ module, only $\Tor^0=\otimes$ and $\Tor^1$ can be
nonvanishing. 

A quick way to argue that $\Tor^1$ always vanishes is the following:
There cannot be any nontrivial differential after $E_2$ in the
\Kunneth{} spectral sequence, so any nonvanishing $\Tor^1$ would
increase the order of the charge groups. But we know from comparison
with the generalized Rosenberg spectral sequence already that $\Tor^0$
accounts for all elements of the \Kgroup{}. 

Nevertheless it would be nice to see directly that $\Tor^1$ has to
vanish. This will be the topic of the remainder of this section.  In
the simpler \BStwist--twisted case we can straightforwardly determine
the derived tensor product (as in~\cite{VolkerLieK}) and find
\begin{equation}
  \label{eq:minustwistTor1}
  \Tor^1_{R(SU_2)}\Big(
  \tKop{(+,\kappa)}_{SU_2}^1\big(SO_3\big),~ \Z
  \Big) = 0 \,,
  \qquad
  \Tor^1_{R(SU_2)}\Big( 
  \tKop{(+,\kappa)}_{SU_2}^0\big(SO_3\big),~ \Z 
  \Big) = 0 \,.
\end{equation}
The \GGtwist--twisted case is more complicated since there are additional
relations, see eq.~\eqref{eq:tKSU2SO3result}. We again have to
distinguish odd and even $\kappa$. 

If $\kappa$ is even, then the \Kgroups{} fit into a short exact
sequence (recall that the polynomials $p_n$ have no constant term):
\begin{equation}
  \vcenter{\xymatrix{
      0 
      \ar[r] & 
      \Z[\Lambda] \left/ 
        { \textstyle
          \left\langle 
            \frac{p_{\kappa}(\Lambda)}{\Lambda}
          \right\rangle 
        } \right. 
      \ar[r]^-{\cdot\Lambda} &
      \tKop{(-,\kappa~\text{odd})}^1_{SU_2}(SO_3)
      \ar[r]^-{\Lambda\mapsto 0} &      
      \Z[\sigma] \Big/\! \langle \sigma^2-1 \rangle 
      \ar[r] &
      0
    }} \,.
\end{equation}
The long exact sequence for $\Tor$ then yields the desired result.

Finally, if $\kappa$ is even then we can use the relation
\begin{equation}
  p_{\kappa}(\Lambda) + (-1)^{\frac{\kappa}{2}}(1+\sigma) 
  = 0 
  \quad \Leftrightarrow \quad
  \sigma = 
  - (-1)^{\frac{\kappa}{2}}  p_{\kappa}(\Lambda) 
  - 1 \,,
\end{equation}
to eliminate $\sigma$ and write 
\begin{equation}
  \tKop{(-,\kappa~\text{even})}^1_{SU_2}(SO_3)
  = \Z[\Lambda] \Big/ \langle \tilde{p}_\kappa(\Lambda) \rangle
  , \qquad
  \tilde{p}_\kappa(2)\not= 0 \,,
\end{equation}
and again we see that the $\Tor^1$ vanishes.

For reference, we get
\begin{equation}
  \label{eq:tKresult}
  \begin{array}{r@{\,=\,}l@{\hspace{15mm}}r@{\,=\,}l}
  \tKop{(+,\kappa)}^1(SO_3) 
  & \Z_\kappa &
  \tKop{(+,\kappa)}^0(SO_3) 
  & \Z_2 \\[1ex]
  \tKop{(-,\kappa)}^1(SO_3) 
  &
  \begin{cases}
    \Z_2\oplus \Z_2 & \kappa~\text{odd} \\
    \Z_4   & \kappa~\text{even}
  \end{cases}
  &
  \tKop{(-,\kappa)}^0 (SO_3)
  & 0 \,.
  \end{array}
\end{equation}
Note that since $\kappa=k+1 = k_\text{GG}/2+1$, this agrees precisely with
the result of~\cite{GaGaSO}. 


\subsection{Comparison with the CFT computation}
\label{sec:gaga}

How does all this relate to the CFT charge computation? We actually
did something very similar. First, note that the twisted equivariant
\Kgroup{}
\begin{equation}
  \label{eq:tKgaga}
  \tKop{(-,\kappa)}^{SU_2}_0\big(SO_3\big) =
  \Big(
  R(SU_2) \oplus 
  \tRop{+}\big(\Z_2 \ltimes U_1\big)  
  \Big)
  \Big/ \Ind\big( R(U_1) \big) \,,  
\end{equation}
is the same $R(SU_2)$ module as the charges $q_0,\dots,
q_{\kappa-2},q_+,q_-$ of Gaberdiel and Gannon, see~\cite{GaGaSO}
eq.~(2.23) --- of course up to our more rational labeling of the
level, i.e. their $n_\text{GG}=\frac{k_\text{GG}}{2}+2=\kappa+1$. To
see this define
\begin{equation}
  \begin{gathered}
    q_\ell \eqdef \Sym^{\ell}(\Lambda) \oplus 0 
    \qquad \forall 0\leq \ell \leq \kappa-2 \\
    q_+ \eqdef 0 \oplus (-1)
    \qquad
    q_- \eqdef 0 \oplus (-\sigma) \,,
  \end{gathered} 
\end{equation}
and take the following $R(SU_2)$ generators for the image of the
Dirac induction:
\begin{subequations}
\begin{align}
  \Ind(\alpha^{-1}) =&~ 
  \Sym^{\kappa-1}(\Lambda) \oplus (1+\sigma)
  \\
  \Ind(\alpha^{-2}) =&~ 
  \Sym^{\kappa-2}(\Lambda) \oplus \Lambda \,.
\end{align}  
\end{subequations}
Then clearly $R(SU_2)$ acts as follows on the $q$ generators:
\begin{equation}
  \label{eq:GaGaqaction}
  \begin{split}
    \Lambda q_0 =&~ \Lambda \Big( 1 \oplus 0 \Big) = q_1
    \\
    \Lambda q_\ell =&~ 
      \Lambda \Sym^\ell(\Lambda) \oplus 0 =
      \Big( \Sym^{\ell-1}(\Lambda)+\Sym^{\ell+1}(\Lambda) \Big)
        \oplus 0 
      = \\ =&~        
      q_{\ell-1} + q_{\ell+1}
      \qquad \forall 1\leq \ell \leq \kappa-3
    \\
    \Lambda q_{\kappa-2} =&~
      \Big( \Sym^{\kappa-3}(\Lambda)+\Sym^{\kappa-1}(\Lambda) \Big)
        \oplus 0 
      = \\ =&~        
      q_{\kappa-3} - \Big( 0 \oplus (1+\sigma) \Big) =
      q_{\kappa-3} + q_+ + q_-
    \\
    \Lambda q_+ =&~ 0 \oplus (-\Lambda) = 
    \Sym^{\kappa-2}(\Lambda) \oplus 0 = q_{\kappa-2}
    \\
    \Lambda q_- =&~ 0 \oplus (-\Lambda) = 
    \Sym^{\kappa-2}(\Lambda) \oplus 0 = q_{\kappa-2} \,.
  \end{split}
\end{equation}
Since Gaberdiel and Gannon were then computing the tensor product of
the $q$ with $\Z$ it is no wonder that they obtain the same result
as we did. However the presentation of the module we are working with,
eq.~\eqref{eq:tKSU2SO3result}, is more useful for the computation of
the tensor product, which is now just one line.


\section{Conclusions and outlook}
\label{sec:conclusions}

In this paper we computed the charges of D--branes on $SO_3$ as
twisted K--theories and found perfect agreement with the CFT results
of~\cite{GaGaSO}. Furthermore, the twisted \Ktheory{} point of view
elucidated certain aspects of supersymmetric\footnote{\Ktheory{} would
  not be relevant for boundary states in the purely bosonic $SO_3$ WZW
  model, as explained in the introduction.} WZW models on non-simply
connected groups, in particular it forces us to study two inequivalent
such theories, which we call \BStwist{} and \GGtwist--twisted $SO_3$
WZW model. The latter is simply the well-known bosonic $SO_3$ WZW
model tensored with free fermions, whereas the former is a novel
theory, that to our knowledge has not been discussed in the literature
so far.

There are two key points in our analysis, which we should emphasize.
Firstly, the twisted K--theories for $SO_3$ come in two guises,
distinguished by a sign $\pm\in H^1(SO_3;\Z_2)=\Z_2$, by which the
K--theory is twisted in addition to the standard twist class
$\kappa\in H^3(SO_3; \Z)$. A very subtle point in the supersymmetric
CFT construction is that the conceptually simpler \GGtwist--twisted
WZW model corresponds to the $- \in H^1(SO_3;\Z_2)$ twisted \Ktheory.
Be that as it may, by applying similar technology as in the work of
Freed, Hopkins and Teleman~\cite{FreedVerlindeAlg, FreedICM,
  FHTcomplex, FHTintegral} we are able to determine the relevant
\Kgroups{} as in~\cite{VolkerLieK, SakuraEquivariant, Sakuratocome}.

Secondly, the additional twisting in the K--theory enforces that there
should be two inequivalent ${\cal N}=1$ supersymmetric WZW models on
$SO_3$, which are precisely distinguished by the sign in $H^1(SO_3;
\Z_2)$. This motivates us to define the \BStwist{} and \GGtwist--twisted
models as a simple current extension of the supersymmetric $SU_2$ WZW
model: there are two simple currents, which have the same action on
the bosonic $\suhat(2)$ part but differ in their action on the free
fermions in $\sohat(3)_1$. The non-trivial action on the fermions
essentially amounts to including $(-1)^F$ in the orbifold action.
This is in particular in accord with the identification of
$H^1(G;\Z_2)$ twisted \Ktheory{} with Hopkins'
$K_{\pm}$~\cite{AtiyahHopkinsKpm}.

There are several avenues in which to extend the present work.
Clearly, it would be very interesting to discuss the novel
\BStwist--twisted $SO_3$ WZW model in more detail, both from the bulk and
boundary CFT point of view. In fact, the construction of the
\BStwist--twisted model obviously applies to the more general setting of
any non-simply connected Lie group $G$ with $|\pi_1(G)|$ even.  The
inequivalent supersymmetric WZW models for $G$ can be obtained by an
analogous simple current construction.

The generalized CFT construction suggested above should of course be
complemented by the corresponding twisted K--theory calculation. We
have seen a beautiful match between the two sides in the $SO_3$ case,
which should generalize to all compact Lie groups. We shall address this
question elsewhere. 
Extensions to orientifolds are also conceivable, e.g. the D--branes
for $SO_3$ have been computed in~\cite{CouchoudSO3orientifold} and the
relevant \Ktheory{} would be a suitably twisted version of Real
\Ktheory.

The key point of this paper is that we cannot simply analyze
supersymmetric WZW models by looking at the bosonic part and
completely ignoring the fermions. A possibly very interesting
application of this would be to revisit the construction of symmetry
breaking boundary conditions in WZW models. Let us sketch the idea:
the analysis of boundary states in WZW models and the corresponding
charge computation has so far essentially neglected the fermions. In
particular, the symmetry-breaking boundary conditions that have been
constructed so far only break parts of the bosonic chiral algebra. It
is however conceivable that new boundary states arise if the free
fermion part of the chiral algebra is partially broken as well, e.g.
by implementing a simple-current action of $j=(-1)^F$ in addition to
the simple currents of the bosonic chiral algebra. 

Recently~\cite{GaGaRo} there has been some progress for $SU(n)$
WZW--models in understanding
the long-standing mismatch between twisted K--theory and WZW
D--brane charges for higher rank simply-connected groups. 
The construction we suggest above gives rise to a set of new boundary
states and it would be important to see whether these
yield additional charges. The proof of completeness
and linear-independence of charges
is certainly one of the main unresolved issue in the CFT analysis of D--brane
charges, a rigorous treatment of which may be inspired by K--theory.


\section*{Acknowledgments}
\label{sec:acknowledgements}

We thank Stefan Fredenhagen for useful discussions and comments on the
manuscript. SSN thanks the University of Pennsylvania for hospitality,
while related work was begun. The work of VB is supported in part by
the NSF Focused Research Grant DMS~0139799 the DOE under contract
No.~DE-AC02-76-ER-03071.


\appendix

\bibliographystyle{JHEP} \renewcommand{\refname}{Bibliography}
\addcontentsline{toc}{section}{Bibliography} 
\bibliography{main}

\begin{thebibliography}{10}
\bibitem{Sakuratocome}
S.~Sch{\"a}fer-Nameki, ``The {$N=2$} chiral ring is twisted equivariant
  {K}--theory.'' To appear.
\end{thebibliography}

\end{document}